\documentclass[preprint,sn-mathphys-num]{sn-jnl}% Math and Physical Sciences Numbered Reference Style
\usepackage{graphicx}%
\usepackage{multirow}%
\usepackage{amsmath,amssymb,amsfonts}%
\usepackage{amsthm}%
\usepackage{mathrsfs}%
\usepackage[title]{appendix}%
\usepackage{xcolor}%
\usepackage{textcomp}%
\usepackage{manyfoot}%
\usepackage{booktabs}%
\usepackage{algorithm}%
\usepackage{algorithmicx}%
\usepackage{algpseudocode}%
\usepackage{listings}%
\theoremstyle{thmstyleone}%
\theoremstyle{thmstyletwo}%

\theoremstyle{thmstylethree}%

\begin{document}

\title[Article Title]{Leapfrogging of a deterministic model for
microeconomic systems in competitive markets}


\author*[1]{\fnm{Alain M.} \sur{Dikand\'e}}\email{dikande.alain@ubuea.cm}

\author[2]{\fnm{H. Ntahombagana} \sur{Matabaro}}\email{hormisdas@aims.edu.gh}
\equalcont{This author contributed equally to this work.}

\affil*[1]{\orgdiv{Laboratory of Research on Advanced Materials and Nonlinear Science, Department of Physics}, \orgname{Faculty of Science, University of Buea},
\postcode{PO Box 63}, \city{Buea},   \country{Cameroon}}

\affil[2]{\orgdiv{African Institute for Mathematical Sciences (AIMS)}, \orgname{Summerhill Estates, East Legon Hills, Santoe}, \city{Accra}, \country{Ghana}}


\abstract{The Behrens-Feichtinger model provides a deterministic picture for the co-evolution of sales of two firms, producing the same goods and competing in a common market. The model involves an active investment strategy such that the temporary investment of each of the two firms depends on its relative position in the market. In this work we are interested in a specific regime of evolution referred to as leapfrogging regime, in which each firm has the possibility to dominate the market alternately during some finite period of time. We examine conditions favoring the leapfrogging dynamics of the model by introducing two appropriate variables, namely the sale difference and total sale of the two firms at any time. An analysis of stability of fixed points of the resulting coupled discrete nonlinear equations is carried out, and the bifurcation diagrams of the sale difference and sum with respect to the elasticity coefficient, are generated. A time-series analysis suggests that the leapfrogging regime, characterized by periodic oscillations of the sale difference from positive to negative branches, is stabilized by specific values of characteristic parameters of the model and when in addition the elasticity coefficient, related to the difference in investment strategies of the two firms, is reasonably high. A positive sign of cumulated sales at any time is required to ensure the availability of goods during the leapfrogging dynamics.}




\maketitle

\section{Introduction}\label{sec1}
\begin{figure}\centering
\includegraphics[width=3in,height= 2.5in]{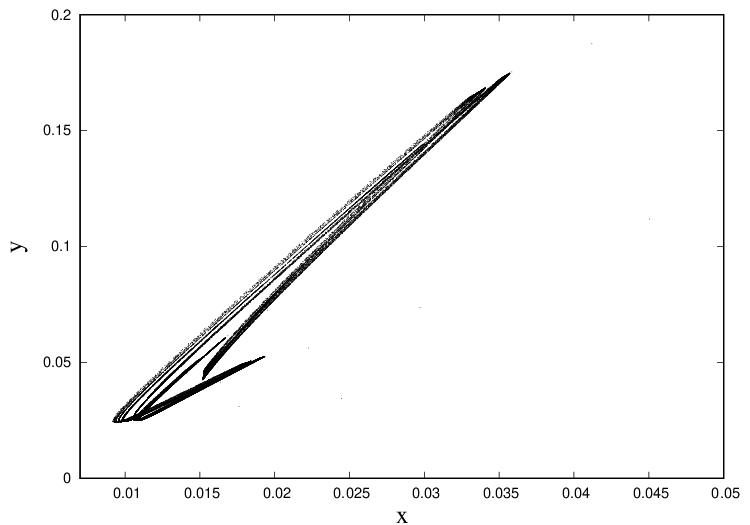}
\caption{Chaotic attractors in phase plane of the coupled discrete mapping (\ref{eq1a})-(\ref{eq1b}), for $a = 0.16$, $b = 0.9$, $\alpha = 0.46$ and $\beta = 0.7$. The observed chaotic orbits have been linked with an existing unstable fixed point for the same parameter values (see e.g. ref. \cite{r10}).}
		\label{fig1}
	\end{figure}

\begin{figure}\centering
\begin{minipage}[c]{0.35\textwidth}
\includegraphics[width=1.75in,height= 1.5in]{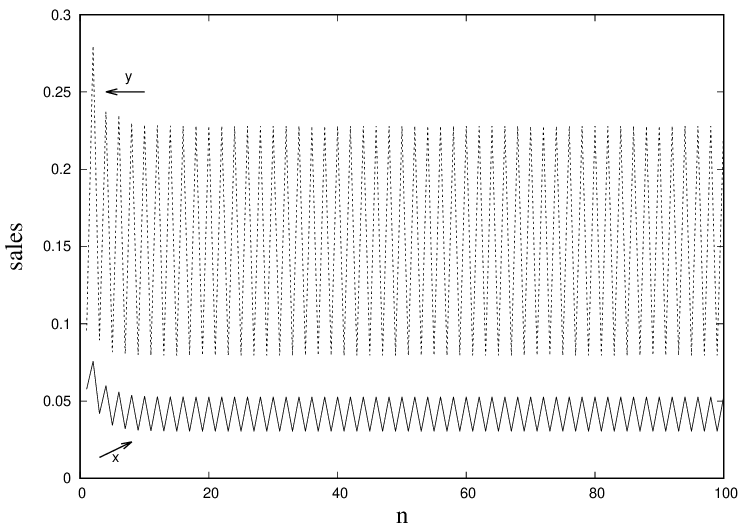}
\end{minipage}%
\begin{minipage}[c]{0.35\textwidth}
\includegraphics[width=1.75in,height= 1.5in]{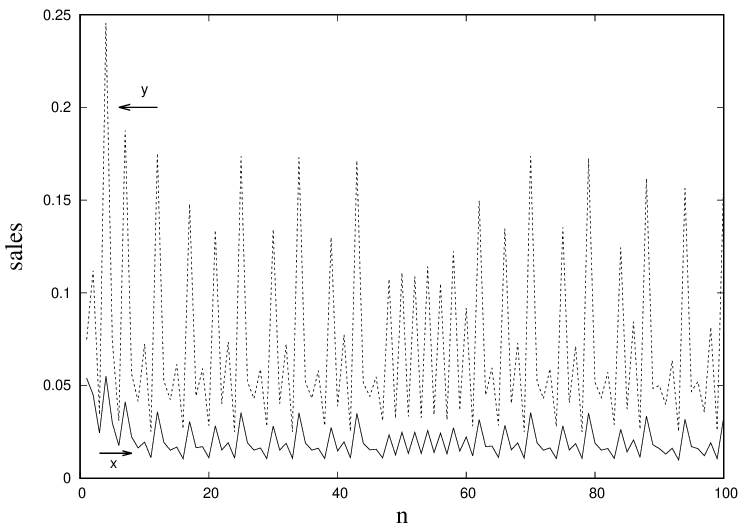}
\end{minipage}%
\begin{minipage}[c]{0.35\textwidth}
\includegraphics[width=1.75in,height= 1.5in]{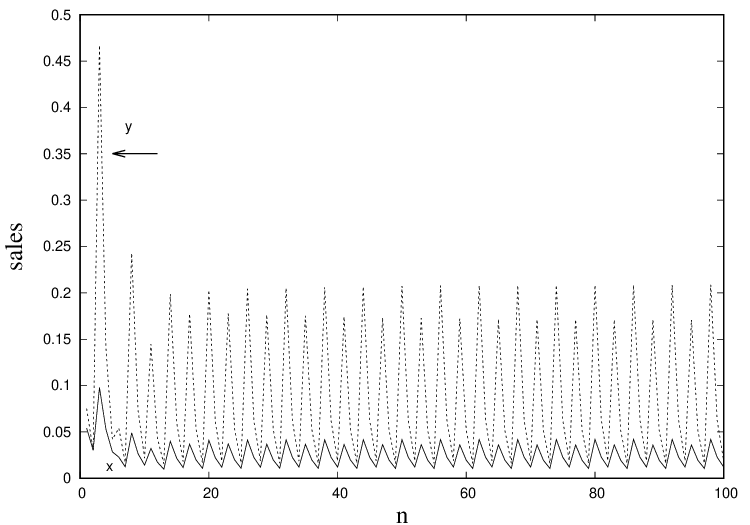}
\end{minipage}
\caption{Time series of sales $x_n$ and $y_n$ of firms $X$ and $Y$ respectively, for $a = 0.16$, $b = 0.9$, $\alpha = 0.46$ and $\beta = 0.7$. The three graphs correspond to three different values of $c$: $c=25$ (left graph), $c=105$ (middle graph) and $c=200$ (right graph).}
\label{fig2}
\end{figure}

Spatiotemporal organizations of modern economies exhibit characteristic features reminiscent of what is commonly known as nonlinear dynamical systems \cite{r1,r1a}. To this last point, over the recent years there has been growing evidences that the extreme complexity of the spatial and temporal evolutions of today's financial systems could be captured by using mathematical models proper to nonlinear dynamical systems' theory \cite{r1a,r2,r3}. Of these models those involving non-randomly driven evolutions of systems, also referred to as deterministic models \cite{r3a,r3b}, have attracted a great deal of interest because of the rich variety of dynamical regimes they can display among which regular and chaotic phases, despite the absence of intrinsic randomness in the systems \cite{r2,r3,r4,r5,r6,r7}. The Behrens-Feichtinger two-partner model \cite{r8,r9,r10,r11,r12,r12a}, which describes the dynamics of two microeconomic systems (for simplicity we shall consider two firms) competing in a common market of goods, is a prototype of such deterministic model. The model assumes that the two firms produce the same goods by adopting asymmetric investment strategies, represented in the model by an "elasticity" parameter that determines how much closer or far aways their respective scales of investment are. More explicitely their temporary investments depend on their relative position on the market, so firm $X$ will invest more when it has an advantage over firm $Y$ whereas firm $Y$ will invest more if it is in a disadvantageous position relative to firm $X$. Denoting by $x_n$ and $y_n$ the relative sales of firms $X$ and $Y$ respectively at the same time $n$ on a discrete time scale, the Behrens-Feichtinger model is represented by the following two coupled logistic-type nonlinear difference equations \cite{r8,r10,r11,r12,r12a}:
\begin{eqnarray}
x_{n+1}&=&(1-\alpha)\, x_n + \frac{a}{1+\exp\big[-c\big(x_n - y_n\big)\big]}, \label{eq1a} \\
y_{n+1}&=&(1-\beta)\, y_n + \frac{b}{1+\exp\big[-c \big(x_n - y_n\big)\big]}. \label{eq1b}
\end{eqnarray}
In the above set the real and positive parameters $\alpha$ and $\beta$ are the rates of decay of sales of firms $X$ and $Y$ respectively, at the discrete time $n$ in the absence of investment. The real parameters $a$ and $b$ are the investment scales of firms $X$ and $Y$ respectively, while the real and positive parameter $c$, most commonly designated "elasticity" coefficient \cite{r10}, is the measure of the extent of the difference in investment strategies of the two firms. Peculiar features of the dynamics of the Behrens-Feichtinger model have been unveiled in several studies \cite{r8,r10,r11,r12}, through time-series analyses or by regarding the coupled set (\ref{eq1a})-(\ref{eq1b}) as a two-dimensional discrete mapping. These studies reveal that solutions to this coupled system are either regular periodic and multi-periodic anharmonic oscillations, or chaotic trajectories depending on specific values of the model parameters  $\alpha$, $\beta$, $a$, $b$ and $c$. Namely a chaotic attractor has been predicted (see fig. \ref{fig1}) for $a=0.16$, $b = 0.9$, $\alpha=0.46$, $\beta=0.7$ and $c=105$ with an unstable fixed point $(x= 0.0118222$, $y=0.0436998)$.
Time-series analysis of solutions $x_n$, $y_n$ are either one-period or multi-periodic anharmonic oscillations depending on values of these parameters. Such periodic anharmonic oscillations, some of which are shown in fig. \ref{fig2}, correspond to dynamical regimes in which sales of the two firms are both regular.
However one of the two firms may decide to increase its gains respective to the other and consequently adopt a truculent investment strategy, resulting in a situation whereby sales of the two firms are no more regularly synchonized, leading to a chaotic regime of evolution of their sales. \\
Apart from the regular and chaotic regimes that have been widely discussed in the literature \cite{r8,r10,r11,r12}, one may also consider a regime in which the investment strategies of the two firms favor alternate controls of the market over finite periods of time. In this specific regime sales of the two firms are expected to remain finite but evolving oppositely over some finite time, such that during this specific period of time sales of one of the two firms are lowest while sales of the other firm are highest. This regime of evolution, in which sales of the two firms "leapfrog" one relative to the other, can turn out to be relevant in a context where the firms compete without impairing their mutual existence and long-term stability in their common market. \\
In this work we investigate conditions for possible emergence of a leapfrogging regime of evolution for the Behrens-Feichtinger model. To this end we introduce two appropriate variables, and use them to reformulate the Behrens-Feichtinger model now as a system of two coupled discrete nonlinear equations descibing the evolutions of the difference and sum of sales of firms $X$ and $Y$. In sec. \ref{sec2} we reformulate the Behrens-Feichtinger model using two new variables, and look for fixed-point solutions of the new system of coupled nonlinear discrete equations. Since the secular equations for these fixed points are not analytically tractable, we use Brent's algorithm to extract the two fixed-points numerically with the elasticity coefficient $c$ used as control parameter. A global stability analysis of the unstable fixed point will also be carried by examining, via the bifurcation diagrams of the sale difference and sum, how the fixed point evolves with increasing $c$. Our aim in so doing is to identify the range of values of $c$ in which the leapfrogging regime is likely to be favored and stabilized. In sec. \ref{sec3} the new set of two coupled discrete nonlinear equations will be solved numerically, and discrete time series of the sale difference and sum will be generated for some specific combinations of values of the model parameters. In these time series the leapfrogging regime will manifest itself via periodic oscillations of the sale difference from positive to negative branches, with the cumulated (or the sum of) sales of the two firms remaining positive at any time $n$. The work ends with a summary and concluging remarks (sec. \ref{sec4}).

\section{Leapfrogging equations: fixed points and bifurcations}\label{sec2}
Consider a specific regime of evolution of the discrete system (\ref{eq1a})-(\ref{eq1b}), favoring a situation whereby the two firms $X$ and $Y$ alternately dominate their common market. In this regime the difference of sales $x_n$ and $y_n$ of the two firms can change periodically from positive to negative values, in a typical evolution reminiscent of leapfrogging familiar in many physical systems \cite{leap1,leap2,leap3,leap4,leap6}. To investigate the dynamics of the discrete system (\ref{eq1a})-(\ref{eq1b}) in the leapfrogging regime, we introduce two new variables i.e. $z_n$ and $w_n$ corresponding respectively to the difference and sum of sales $x_n$ and $y_n$ at time $n$:
\begin{equation}
 z_n= x_n - y_n, \qquad  w_n= x_n + y_n. \label{nv}
\end{equation}
With these two new variables, equations (\ref{eq1a}) and (\ref{eq1b}) become:
\begin{eqnarray}
z_{n+1}&=&(1-\alpha_1)\, z_n + \beta_1\,w_n- \frac{b_1}{1+\exp\left(-c z_n\right)}, \label{eq6a} \\
w_{n+1}&=&(1-\alpha_1)\, w_n + \beta_1\,z_n + \frac{a_1}{1+\exp\left(-c z_n\right)}, \label{eq6b}
\end{eqnarray}
where:
\begin{equation}
\alpha_1=\frac{\alpha +\beta}{2}, \qquad \beta_1=\frac{\beta-\alpha}{2}, \qquad
a_1= a+b, \qquad b_1=b-a. \label{f}
\end{equation}
The new system of discrete equations given in formula (\ref{eq6a})-(\ref{eq6b}) is also nonlinear, and according to formula (\ref{f}) characteristic parameters of these coupled discrete nonlinear equations are determined by parameters of the Behrens-Feichtinger model (\ref{eq1a})-{\ref{eq1b}). \\
Fixed points of the coupled system eqs. (\ref{eq6a})-(\ref{eq6b}) are pairs $(z_0, w_0)$ of solutions of the following $2\times 2$ nonlinear algebraic equations:
\begin{eqnarray}
\alpha_1 z_0 - \beta_1 w_0 + \frac{b_1}{1+\exp\left(-c z_0\right)}&=&0, \label{eq7a} \\
\alpha_1 w_0 - \beta_1 z_0 - \frac{a_1}{1+\exp\left(-c z_0\right)}&=&0. \label{eq7b}
\end{eqnarray}
The latter system was solved numerically using Brent's algorithm, and for illuations we considered the same values of characteristic parameters used in the previous curves namely $a = 0.16$, $b = 0.9$, $\alpha = 0.46$ and $\beta = 0.7$. The elasticity coefficient $c$, which measures the difference in investment strategies of the two firms, stands for a suitable control parameter and will serve as variable for the numerically generated fixed points. In fig. \ref{fig3} we plot $z_0$ (line) and $w_0$ (dots) versus $c$. We see that when $c=0$, $z_0$ takes a negative threshold value and grows asymptotically toward zero as $c$ increases. Concomitantly $w_0$ is at a positive threshold for $c=0$, decreasing asymptotically toward zero as $c$ becomes very large.
\begin{figure}\centering
\includegraphics[width=3.in, height= 2.5in]{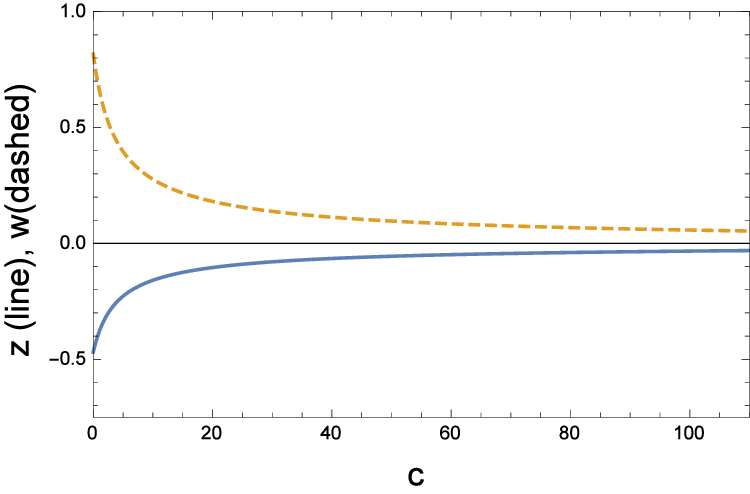}
\caption{Roots $z_0$ (full line) and $w_0$ (dashed line) of the coupled system (\ref{eq7a})-(\ref{eq7b}), plotted versus $c$ for $a = 0.16$, $b = 0.9$, $\alpha = 0.46$ and $\beta = 0.7$.}
		\label{fig3}
	\end{figure}
Note that $z_0$ and $w_0$ are related through a linear relation, as can be shown by combining eqs. (\ref{eq7a}) and (\ref{eq7b}). Indeed from these two formula we obtain:
\begin{equation}
w_0= z_0\,\frac{a_1\alpha_1 - b_1\beta_1}{a_1\beta_1 - b_1\alpha_1}. \label{eq8}
\end{equation}
This relation suggests a choice of paramters in such a way that $a_1\beta_1-b_1\alpha_1 \neq 0$. Moreover, the variations of $z_0$ and $w_0$ observed in fig. \ref{fig3} indicate that the sum of sales $w_0$ will always be positive and the sale difference $z_0$ always negative, for the specific values chosen for characteristic parameters of the model.\\
Let us examine the stability of the two-dimensional discrete mapping (\ref{eq6a})-(\ref{eq6b}), against infinitesimal deviations $(\delta z_{n+1},\, \delta w_{n+1})$ with respect to the fixed point $(z_0, w_0)$. The equations accounting for these infinitesimal transformations can be represented by the linear $2\times 2$ matrix equation:
\begin{equation}
(\delta z_{n+1},\, \delta w_{n+1})= J_c(z_0, w_0)\,(\delta z_n,\, \delta w_n), \label{a11}
\end{equation}
where $J_c(z_0, w_0)$, the Jordan matrix governing the infinitesimal transfomations with respect to the fixed point $(z_0, w_0)$, is obtained as:
\begin{eqnarray}
J_c(z_0, w_0) = \left(\begin{array}{cc}
1 -\alpha_1 - \frac{c\, b_1}{4\cosh^2(c\,z_0/2)} & \beta_1 \\
\beta_1 +  \frac{c\,a_1}{4\cosh^2(c\,z_0/2)} & 1-\alpha_1  \end{array} \right). \label{a12}
\end{eqnarray}
Period-one orbits of the two-dimensional mapping are points $(z, w)$ generated by successive iterates from the fixed points, leading to a sequence $(z_0, w_0)$, $(z_1, w_1)$, $\cdots$, $(z_N, w_N)$ via the linear transformation (\ref{a11}). In this sequence, stable period-one orbits will require $\vert\, trace [J_c(z_0, w_0)]\, \vert \leq 2$ suggesting the following stability condition for period-one orbits:
\begin{equation}
8(1 -\alpha_1) - c\, b_1\,sech^2(c\,z_0/2)=\pm 8. \label{cda}
\end{equation}
For the fixed point $(z_0,w_0)=(0, 0)$ emerging from fig. \ref{fig3} for large $c$, the requirement (\ref{cda}) leads to two possible values of $c$: a negative value which is irrelevant and therefore should be ruled out, and a positive value expressed in terms of characteristic parameters of the model as:
\begin{equation}
c_c=\frac{16 + 8\alpha_1}{b_1}.  \label{cdb}
\end{equation}
Since the quantity $c_c$ obtained in eq. (\ref{cdb}) is the critical value of the elasticity parameter $c$ for which period-one orbits are unstable, it is ready to expect a period-doubling bifurcation at $c=c_c$. The bifurcations diagrams with respect to the elasticity coefficient $c$, of the difference $z$ (left graph) and sum $w$ (right graph) of sales of the two firms, are plotted in fig. \ref{fig4} for $a = 0.16$, $b = 0.9$, $\alpha = 0.46$ and $\beta = 0.7$. These parameter values imply $b_1=0.74$ and $\alpha_1=0.58$, such that formula (\ref{cdb}) yields $c_c=27.891891892$. This is consistent with the value of $c$ where the first period-doubling bifurcation is observed in the two bifurcation diagrams shown in fig. \ref{fig4}.
\begin{figure}\centering
\begin{minipage}[c]{0.53\textwidth}
\includegraphics[width=2.5in, height= 2.in]{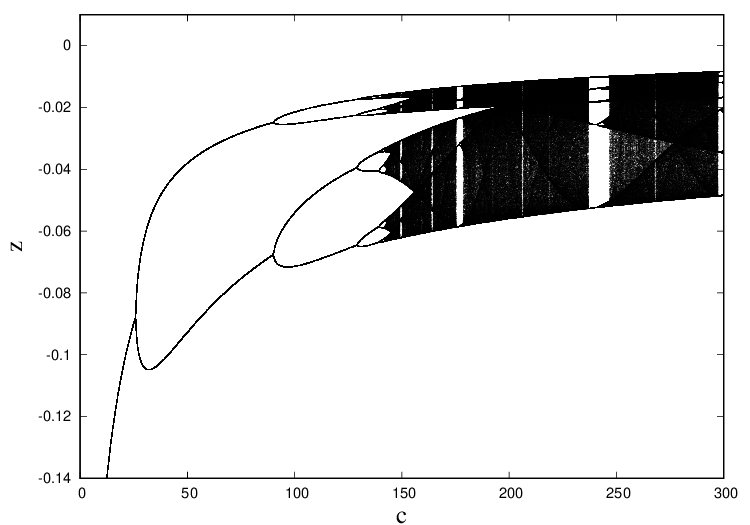}
\end{minipage}%
\begin{minipage}[c]{0.53\textwidth}
\includegraphics[width=2.5in, height= 2.in]{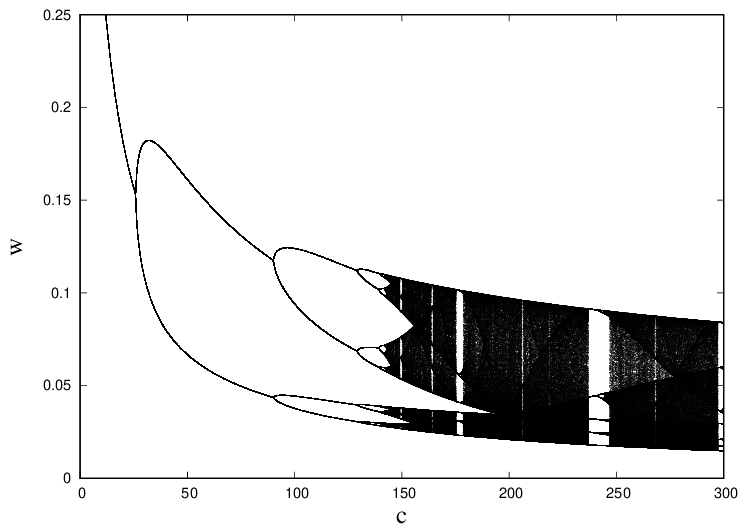}
\end{minipage}
\caption{Bifurcation diagrams with respect to $c$ of the sale difference $z$ (left graph) and sum $w$ (right graph), for $b_1=0.74$ and $\alpha_1=0.58$.}
\label{fig4}
\end{figure}

\section{Time-series analysis of leapfrogging dynamics}\label{sec3}
The fixed-point solution shown in fig. \ref{fig3}, as well as the bifurcation diagrams with respect to $c$ in fig. \ref{fig4}, suggest that the sale difference $z$ is always negative and the sum of sales $w$ always positive, both eventually approaching zero asymptotically as $c$ gets larger and larger. These behaviours are actually not universal, in fact they are specific to values of $a$, $b$, $\alpha$ and $\beta$ used in numerical simulations for their generations. Concretely for the leapfrogging evolution it is desired that $z$ should oscillate periodically from positive to negative values, reflecting alternate leading positions of firms over some finite period of time. The time series of $x_n$ and $y_n$ shown in fig. \ref{fig1} already indicated a long-term monopoly of firm $Y$, in fig. \ref{fig5} we plotted the time series of $z_n$ (full line) and $w$ (dots) for $a = 0.16$, $b = 0.9$, $\alpha = 0.46$, $\beta = 0.7$, and two different values of $c$ selected to highlight the absence of leapfrogging for all orders of magnitude of $c$. As $c$ is increased the sale difference $z_n$ oscillates periodically but remains in the branch of negative values.
\begin{figure}\centering
\begin{minipage}[c]{0.53\textwidth}
\includegraphics[width=2.5in, height= 2.in]{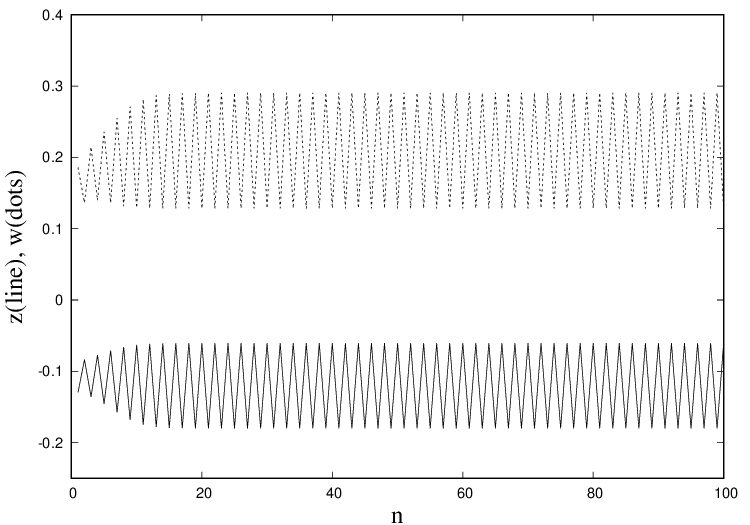}
\end{minipage}%
\begin{minipage}[c]{0.53\textwidth}
\includegraphics[width=2.5in, height= 2.in]{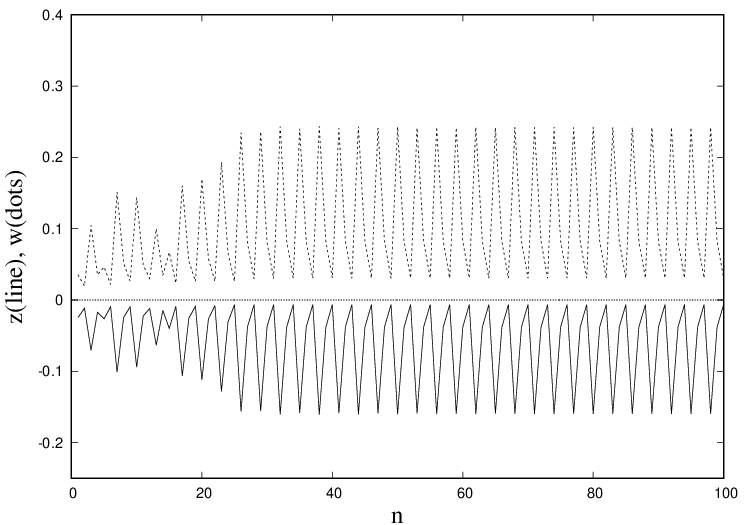}
\end{minipage}
\caption{Discrete time series of sale difference $z_n$ (full line) and sum $w_n$ (dots), for $a=0.16$, $b=0.9$, $\alpha=0.46$ and $\beta=0.7$. Left graph: $c=20$, right graph`: $c=200$.}
\label{fig5}
\end{figure}
In clear the values $a=0.16$, $b=0.9$, $\alpha=0.46$ and $\beta=0.7$, used in the above analysis and in most studies including \cite{r10}, are not relevant for the leapfrogging dynamics of the system. Fig. \ref{fig6} shows plots of time series of $z_n$ and $w_n$ now for $a=0.7$, $b=0.45$, $\alpha=0.7$, $\beta=0.4$ and three different values of $c$ i.e. $c=10$ (left graph), $c=20$ (middle graph) and $c=150$ (right graph). This last set of parameter values favors leapfrogging, and as it is noticeable the leapfrogging dynamics is enhanced by an increase of $c$.
\begin{figure}\centering
\begin{minipage}[c]{0.35\textwidth}
\includegraphics[width=1.7in, height= 1.5in]{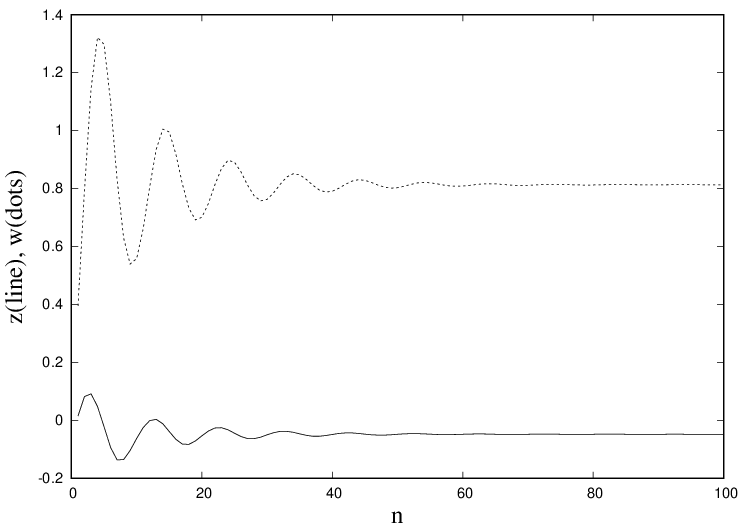}
\end{minipage}%
\begin{minipage}[c]{0.35\textwidth}
\includegraphics[width=1.7in, height= 1.5in]{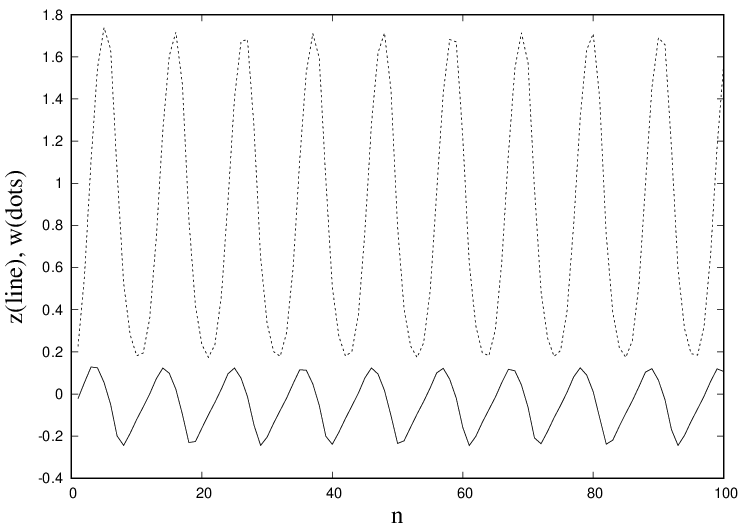}
\end{minipage}%
\begin{minipage}[c]{0.35\textwidth}
\includegraphics[width=1.7in, height= 1.5in]{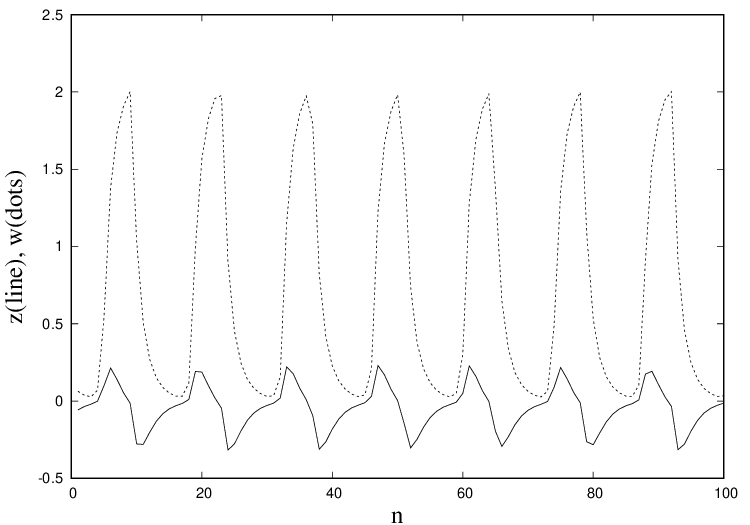}
\end{minipage}%
\caption{Discrete time series of sale difference $z_n$ (full line) and sum $w_n$ (dots), for $a=0.7$, $b=0.45$, $\alpha=0.7$ and $\beta=0.4$. Left graph: $c=10$, middle graph: $c=20$ right graph: $c=150$.}
\label{fig6}
\end{figure}
In fig. \ref{fig7} we plotted $z_n$ and $w_n$ versus $n$ for $a=0.9$, $b=0.6$, $\alpha=0.7$ and $\beta=0.46$, and in fig. \ref{fig8} the same quantities were plotted for $a=0.95$, $b=0.35$, $\alpha=0.85$ and $\beta=0.15$. In these two figures the leapfrogging dynamics is also observed at relatively high values of the elasticity coefficient $c$.
\begin{figure}\centering
\begin{minipage}[c]{0.35\textwidth}
\includegraphics[width=1.7in, height= 1.5in]{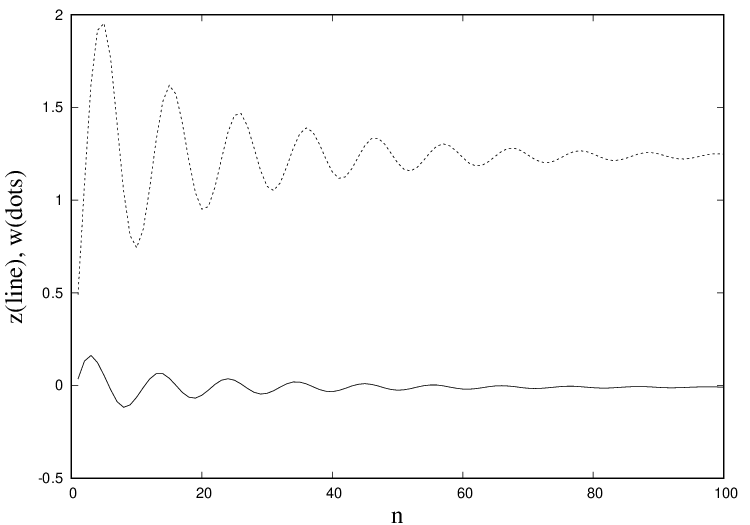}
\end{minipage}%
\begin{minipage}[c]{0.35\textwidth}
\includegraphics[width=1.7in, height= 1.5in]{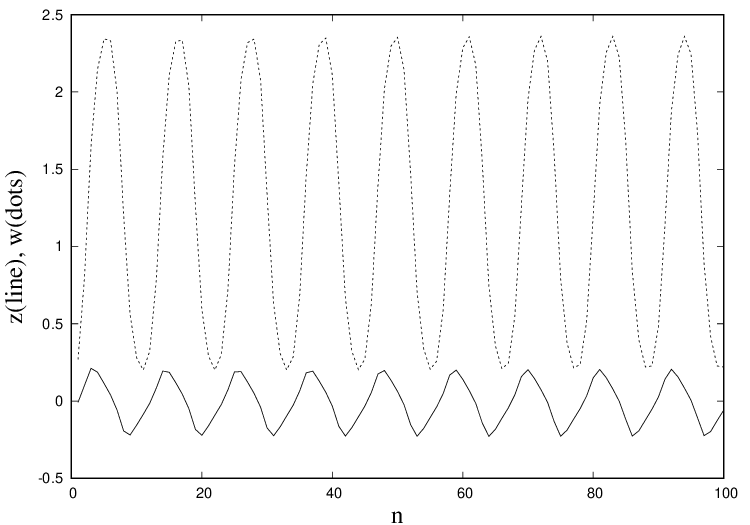}
\end{minipage}%
\begin{minipage}[c]{0.35\textwidth}
\includegraphics[width=1.7in, height= 1.5in]{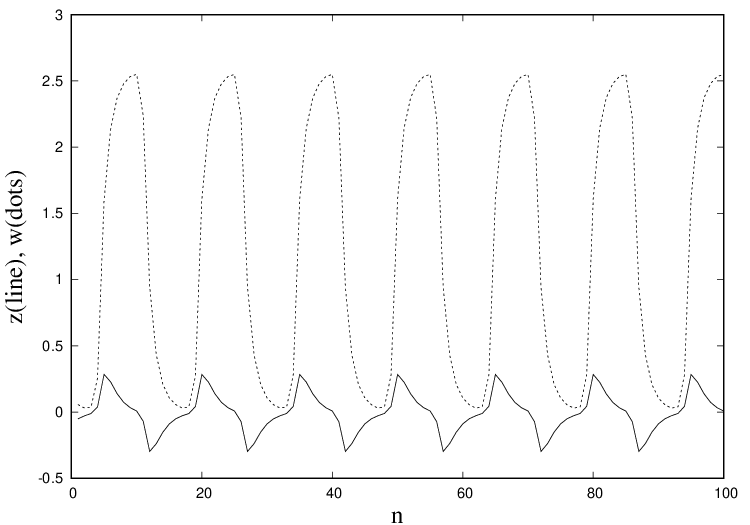}
\end{minipage}%
\caption{Discrete time series  of sale difference $z_n$ (full line) and sum $w_n$ (dots), for $a=0.9$, $b=0.6$, $\alpha=0.7$ and $\beta=0.46$. Left graph: $c=10$, middle graph: $c=20$, right graph: $c=150$.}
\label{fig7}
\end{figure}
\begin{figure}\centering
\begin{minipage}[c]{0.35\textwidth}
\includegraphics[width=1.7in, height= 1.5in]{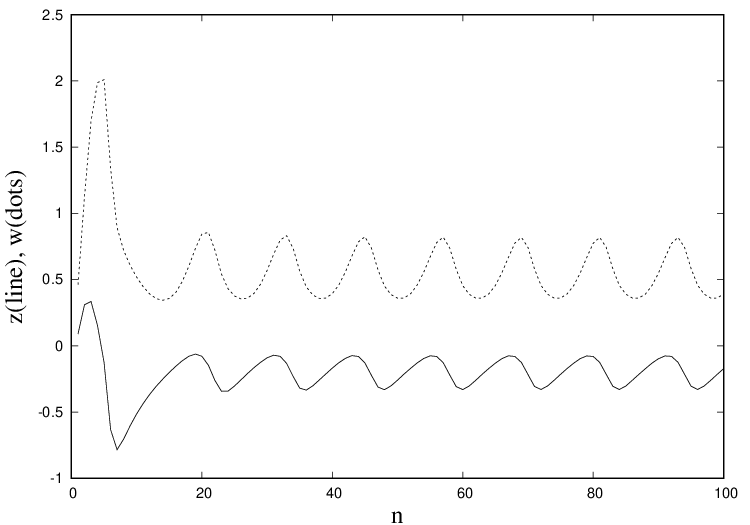}
\end{minipage}%
\begin{minipage}[c]{0.35\textwidth}
\includegraphics[width=1.7in, height= 1.5in]{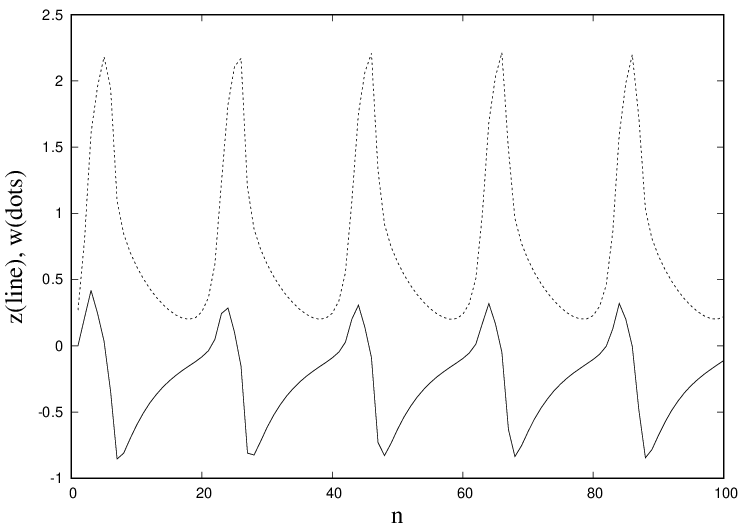}
\end{minipage}%
\begin{minipage}[c]{0.35\textwidth}
\includegraphics[width=1.7in, height= 1.5in]{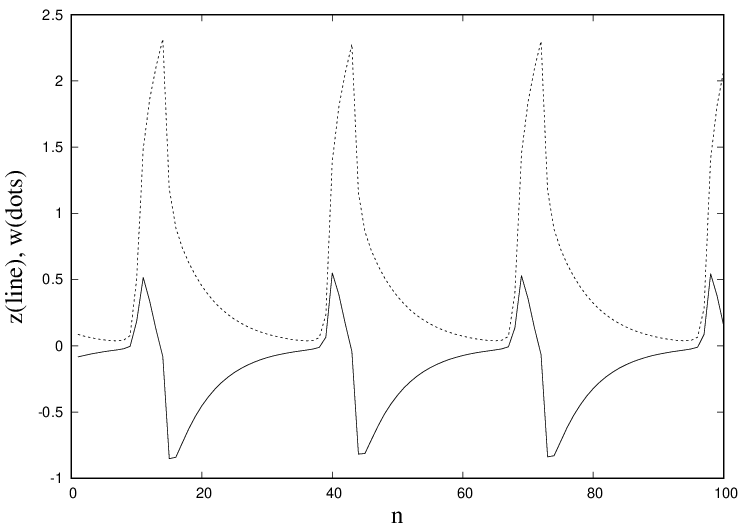}
\end{minipage}%
\caption{Discrete time series  of sale difference $z_n$ (full line) and sum $w_n$ (dots), for $a=0.95$, $b=0.35$, $\alpha=0.85$ and $\beta=0.15$. Left graph: $c=10$, middle graph: $c=20$, right graph: $c=150$.}
\label{fig8}
\end{figure}
A quite interesting fact about the emergence of leapfrogging from the coupled discrete nonlinear equations (\ref{eq6a}-(\ref{eq6b}), is that the rate of decay of sales of firm $X$ should be greater or at least equal to the corresponding parameter for firm $Y$ (i.e. $\alpha \leq \beta$). In addition the investment scale $a$ of firm $X$ should greater or at least equal to $b$, the scale of investment of firm $Y$ (i.e. $a\leq b$). It is also remarkable that the cumulated instantaneous sale of the two firms i.e. $w_n$ is always positive, eventually approaching zero but never cross toward negative values. The latter behaviour ensures the non-occurence of shortage of goods during the leapfrogging evolution of the two firms.

\section{Summary and concluding remarks}\label{sec4}
The Behrens-Feichtinger model has been considered in a large number of past studies and it is well established that the model possesses a rich spectrum of dynamical properties, ranging from regular periodic anharmonic oscillations to chaotic evolutions depending on values of its characteristic parameters. In the present study we examined the dynamics of the system, focusing on a specific regime of evolution that we referred to as leapfrogging. In this specific regime sales of the two firms are assumed to evolve oppositely, such that in finite periods of time sales of one firm are dominant over sales of the other firm and vice-versa. The leapfrogging dynamics was described by periodic oscillations of a variable related to the difference of sales of the two firms, which is expected to change periodically from positive to negative values as a result of the leapfrogging dynamics. We obtained that this particular regime of evolution requires appropriate choice of values of characteristic parameters of the model. It is relevant that the total sum of sales at any time $n$ is always positive, ensuring the availability of products for the satisfaction of consumers.

\section*{Acknowledgements}

A. M. Dikand\'e acknowledges support from the Alexander von Humboldt foundation, within the framework of the "Return Fellowship program" for a visit at the Max-Planck Institute for the Physics of Complex Systems (MPIPKS). H. N. Matabaro thanks the African Institute for Mathematical Sciences (AIMS) for sponsoring his academic stay at AIMS-Ghana.

\section*{Declarations}

\subsection*{Funding}
No funding was requested for the present work.

\subsection*{Conflict of interest}
The authors certify that they have no affiliations or involvement with any organization or
entity with any financial interest, or non-financial interest in the subject matter or materials discussed in this
manuscript.

\subsection*{Data availability}
Data sharing is not applicable to this article as no datasets were generated or analyzed
during the current study.

\subsection*{Authorship contribution statement}

A. M. Dikand\'e: Conceptualization, Investigation, numerical simulations, validation, writing of the original draft. \\
H. N. Matabaro: Investigation, Formal analysis, validation.


\begin{thebibliography}{99}%

\bibitem{r1} Weidlich and Braun M., "The master equation approach to nonlinear economics", J. Evol. Econ. {\bf  2}, 233-265 (1992).
\bibitem{r1a}Lines M. (Editor), "Nonlinear dynamical systems in Economics" (Springer, Wien-New York, 2005).
\bibitem{r2}Lorenz H. W., "Nonlinear Dynamical Economics and Chaotic Motion" (2nd Edition, Springer-Verlag, New York-Berlin, Heidelberg, 1997).
\bibitem{r3} Chen P., "Empirical and theoretical evidence of economic chaos", Syst. Dynam. Rev. {\bf 4}, 81-108 (1998).
\bibitem{r3a}Sorger R., "Dynamic Economic Analysis: Deterministic Models in Discrete Time" (Cambrige University Press, Cambridge, 2015).
\bibitem{r3b}Dockner E. J., Hartl R. F., Luptacik M. and Sorger G. (Eds.). "Optimization, Dynamics, and Economic Analysis: Essays in Honor of Gustav Feichtinger" (Springer, Berlin, 2000).
\bibitem{r4}Kopel M., "Improving the performance of an economic system: controlling chaos", J. Evol. Econ. {\bf 7} (1997) 269-289.
\bibitem{r5}Puu T., "Attractors, Bifurcation, and Chaos:
Nonlinear Phenomena in Economics" (Springer, New York, 2000).
\bibitem{r6}Orlando G., Bufalo M. and Stoop R., "Financial markets? deterministic aspects modeled by a low-dimensional equation", Scientific Reports {\bf 12}, 1693 (2022).
\bibitem{r7}Demmel R., "The basic deterministic macroeconomic model", in: Fiscal Policy, "Public Debt and the Term Structure of Interest Rates. Lecture Notes in Economics and Mathematical Systems", vol 476. (Springer, Berlin, Heidelberg, 1999).
\bibitem{r8}Feichtinger G., in: Haag G., Mueller U., Troitzsch K. G. (Eds.), "Economic Evolution and Demographic
Change" (Springer, Berlin, 1992).
\bibitem{r9}Holyst J. A., Hagel T., Haag G. and Weidlich W., "How to control a chaotic economy?", J. Evol. Econ. \textbf{6}, 31-42 (1996).
\bibitem{r10}Holyst J. A. and Urbanowicz K., "Chaos control in economical model by
time-delayed feedback method", Physica A {\bf 287}, 587-598 (2000).
\bibitem{r11}Salarieh H. and Alasty A., "Delayed feedback control via minimum entropy strategy in an economic model", Physica A {\bf 387}, 851-860 (2008).
\bibitem{r12}Behrens D. A., Feichtinger G. and Prskawetz A., "Complex dynamics and control of arms race",  Eur. Jour. Oper. Res. \textbf{100}, 192-215 (1997).
\bibitem{r12a}Perc M., "Microeconomic uncertainties facilitate cooperative
alliances and social welfare", Economics Let. \textbf{95}, 104-109 (2007).
\bibitem{leap1}Malomed B. A., "Leapfrogging solitons in a system of coupled KdV equations", Wave Motion {\bf 9}, 401-411 (1987).
\bibitem{leap2}Liu A. K., Pereira N. R., Ko D. R. S, "Weakly interacting internal solitary waves in neighbouring pycnoclines", J. Fluid Mech. {\bf 122}, 187-194 (1982).
\bibitem{leap3}Weidman P. D. and Johnson M., "Experiments on leapfrogging
internal solitary waves", J. Fluid Mech. \textbf{122}, 195-213 (1982).
\bibitem{leap4}Nitsche M., Weidman P. D., Grimshaw R., Ghrist M. and Fornberg B., "Evolution of solitary waves in a two-pycnocline system",
J. Fluid Mech. {\bf 642}, 235-277 (2010).
\bibitem{leap6}Achere Nkongho A., Akong Ngate L., Dikand\'e A. M. and Essimbi B. Z., "Leapfrogging dynamics of interacting solitons in weakly coupled
nonlinear transmission lines", SN Applied Sciences
\textbf{1}, 552 (2019).

\end{thebibliography}
\end{document}